\documentclass[showpacs,amssymb,amsmath,nobibnotes, superscriptaddress
,aip,pof%
,tightenlines
,twocolumn
]{revtex4-1}
\usepackage{bm}
\usepackage{graphicx}

\begin{document}
\textfloatsep 10pt

\title{Robust inverse energy cascade and turbulence structure in three-dimensional layers of fluid}

\author{D. Byrne}
\author{H. Xia}
\email{Hua.Xia@anu.edu.au}
\author{M. Shats}

\affiliation{Research School of Physics and Engineering,\\ The Australian National
University, Canberra ACT 0200, Australia}

\date{\today}


\begin{abstract}
Here we report the first evidence of the inverse energy cascade in a flow dominated by 3D motions. Experiments are performed in thick fluid layers where turbulence is driven electromagnetically. It is shown that if the free surface of the layer is not perturbed, the top part of the layer behaves as quasi-2D and supports the inverse energy cascade regardless of the layer thickness. Well below the surface the cascade survives even in the presence of strong 3D eddies developing when the layer depth exceeds half the forcing scale. In a bounded flow at low bottom dissipation, the inverse energy cascade leads to the generation of a spectral condensate below the free surface. Such coherent flow can destroy 3D eddies in the bulk of the layer and enforce the flow planarity over the entire layer thickness.
\end{abstract}

\maketitle

\section{Introduction}
There has been remarkable progress in the understanding of turbulence in fluid layers. Such layers, characterized by large aspect ratios, are ubiquitous in nature. 2D turbulence theory by Kraichnan \cite{Kraichnan1967}, in particular the inverse energy cascade, has been confirmed in experiments in thin fluid layers \cite{Sommeria1986, Paret_Tabeling1998, Belmonte1999,Vorobieff1999,Chen2006,Shats2005}. More recent studies showed that theoretical results derived for idealized 2D turbulence are valid in a variety of conditions, even when the theory assumptions are violated \cite{Xia2008, Xia2009}.

In bounded turbulence, the inverse energy cascade may lead to the accumulation of spectral energy at the box size scale and the generation of a spectral condensate, a large vortex coherent over the flow domain \cite{Xia2009}. Good agreement with the Kolmogorov-Kraichnan theory was found in the double layers of fluids in spectrally condensed turbulence \cite{Xia2009,Xia2011}.

It should be noted that though spectral condensation was observed in the double-layer configurations \cite{Paret_Tabeling1998, Shats2005, Shats2007, Xia2008, Xia2009}, in single layers of electrolytes not only spectral condensation, but also the very existence of the inverse energy cascade has been questioned. For example, in Ref. \cite{Merrifield2010} the flow generated in a single layer was referred to as ``spatio-temporal chaos'' to stress the absence of the turbulent energy cascades. It is thus important to better understand spatial structure of turbulence in such layers as well as differences between turbulence in double and single layers. (Here we do not discuss MHD flows which also show spectral condensation \cite{Sommeria1986}, but where two-dimensionality is imposed by homogeneous magnetic field and bottom dissipation is restricted to a thin Hartmann layer.)

A comparison of turbulent flows in a single and double-layer configuration is also important to improve our understanding of turbulence in atmospheric boundary layers. These layers are very different over terrain and the oceans, with the former being substantially thicker than the latter ones. Stable immiscible layers of fluids have been generated in the laboratory by placing a heavier non-conducting fluid at the bottom of the cell and a lighter layer of electrolyte resting on top of it \cite{Chen2006, Xia2008, Xia2009}. In this case the electromagnetic forcing is detached from the solid bottom and it is maximal just above the interface between the two fluids. The structure of the flow in the top layer close to the interface with the bottom layer may be similar to that of the atmospheric boundary layer over the ocean.

In this paper we present new results on the spatial structure of turbulence in a single- and a double-layer turbulent flow. In contrast to the previous studies, we focus here on \textit{thick} layers. Recent 3D direct numerical simulations of the Navier-Stokes equations \cite{Celani2010} have shown that the finite layer depth leads to splitting of the energy flux. In thin layers the energy flux injected by forcing is inverse. As the layer thickness $h$ is increased, a larger fraction of the flux is redirected down scale, towards the wave numbers larger than the forcing wave number $k_f$. At $h/l_f > 0.5$, most of the injected flux cascades to small scales.

Experiments in thick single layers of electrolyte partially confirm this picture \cite{Shats2010}. It has been shown that at $h/l_f > 0.5$ strong 3D motions indeed appear in a layer. Such 3D eddies increase vertical flux of the horizontal momentum (eddy viscosity) which leads to a bottom drag dissipation rate higher than the one expected from a quasi-2D model. However, no evidence of the direct energy cascade was found. At $k > k_f$ no $k^{-5/3}$ spectrum was observed as should be expected in the presence of the direct energy cascade. Probably 3D eddies over the solid bottom introduce higher damping for small scales ($k > k_f$) and do not allow the direct cascade to develop.

In the double layers the effects of the solid bottom on the top layer turbulence are isolated by a layer of non-conducting, low viscosity heavier fluid at the bottom. It was recently found\cite{Xia2011} that the range of depths in which a top layer flow remains planar and supports spectral condensation, is greatly extended, $h/l_f > 0.5$. It was hypothesized,  that a residual inverse energy flux condenses turbulent energy into large scale flows even in thick layers. These coherent flows enforce the planarity by shearing off vertical eddies and thus secure the upscale energy transfer \cite{Xia2011}. It is not clear however how the large scale flow is generated in a thick layer in the first place. In this paper we present new results which shed light on this.

\section{Experimental setup and flow characterization}

\begin{figure*}
\centerline{\includegraphics[width=12.0cm]{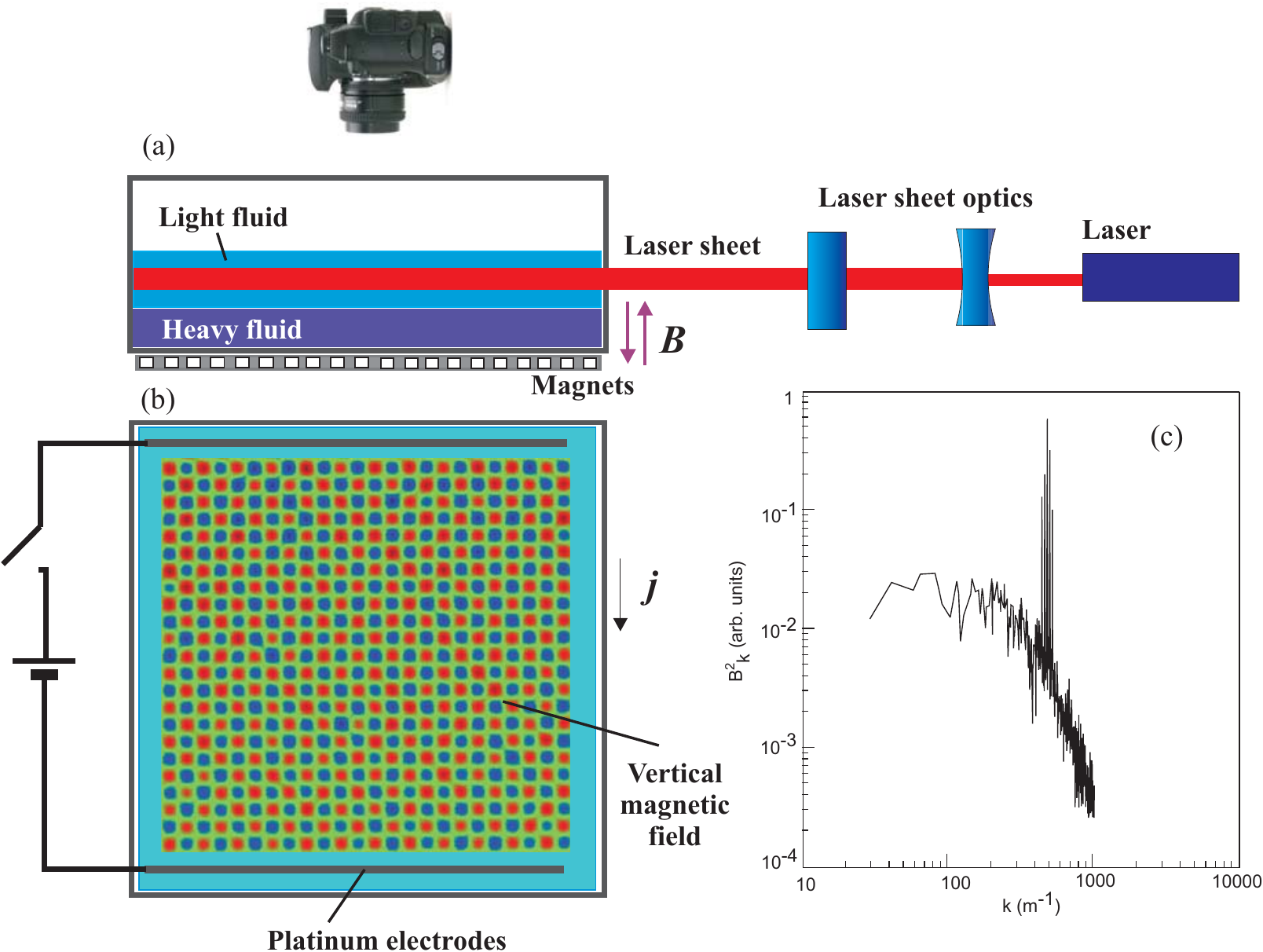}}
\caption{\label{fig1} Schematic of experimental setup. (a) Neutrally buoyant seeding particles in the top (conducting) layer are illuminated and their motion is filmed from above. (b) Measured vertical magnetic field produced by the magnet array: blue and red dots indicate upward and downward $B$ direction. (c) Wave number spectrum of the measured magnetic field.}
\end{figure*}

In these experiments turbulence is generated through the interaction of a large number of electromagnetically driven vortices \cite{Shats2005, Xia2008, Xia2009}. An electric current flowing across the conducting fluid layer interacts with the spatially varying vertical magnetic field produced by a 24 $\times$ 24 or 30 $\times$ 30 array of magnetic dipoles (10 mm and 8 mm separation respectively), Figure~\ref{fig1}. The magnet arrays are placed under the bottom of the 0.3 $\times$ 0.3 m$^2$ fluid cell. To ensure that turbulence is forced monochromatically at $k = k_f$, vertical magnetic field produced by the array has been measured using a Hall probe scanned in horizontal planes at several heights above the array. The measured magnetic field (Fig.~\ref{fig1}(b)) has then been Fourier transformed in 2D, Fig.~\ref{fig1}(c). The spectrum shows that $\mathbf{J} \times \mathbf{B}$ forcing is localized in $k$-space in a narrow spectral range (in this example, for a 10 mm magnet separation, the spectrum peaks at $k \approx 630$  rad$\cdot$m$^{-1}$). The forcing strength is controlled by adjusting electric current through the layer in the range 0.5-5 A.

Turbulence is generated either in a single layer of $Na_2SO_4$ water solution, or it is driven in the top layer of the electrolyte which rests upon a layer of heavier (1820 kg/m$^3$) nonconducting liquid (FC-3283 by 3M) which is not soluble in water. In the latter case, forcing is the strongest just above the interface between the layers.

The flow is visualized using neutrally buoyant seeding particles 50 $\mu$m in diameter illuminated by a horizontal laser sheet. The thickness of the laser sheet and its height relative to the free surface are adjusted to visualize different regions of the layers. Perturbations to the fluid surface are monitored by reflecting a thin laser beam off the free surface onto a distant screen (the detection sensitivity is $\sim 5\times10^{-3}$ mm). In all reported experiments no perturbation to the free surface is detected.

To characterize the vertical structure of the flow, defocusing PIV is used \cite{Willert1992}.  This technique uses a single CCD camera with a multiple pinhole mask to measure three-dimensional velocity components of individual seeding particles in the flow \cite{Shats2010}. The horizontal turbulent velocity fields are derived using particle image velocimetry (PIV). Particle pairs are matched from frame to frame throughout the illuminated volume using a PIV/PTV hybrid algorithm. Derived velocities are then averaged over hundreds of frame pairs to generate converged statistics of the root-mean-square velocities  $\left\langle V_{x,y,z} \right\rangle_{rms}$ throughout the layer. The technique allows one to resolve vertical velocities  $\left\langle V_{z} \right\rangle_{rms} \geq$ 0.4 mm/s  and $\left\langle V_{xy} \right\rangle_{rms} \geq$ 2.5$\times 10^{-2}$ mm/s.  The damping rate of the flow is studied from the decay of the horizontal energy density after switching off the forcing \cite{Boffetta_EPL_2005}.

We also study vertical motion of seeding particles by illuminating the layer using a thick vertical laser sheet. Streaks of the particles in $z-x$ plane are filmed with the exposure time of 1-2 s through a transparent side wall of the fluid cell.

\section{Turbulent flow in a thick single layer}

\begin{figure}
\centerline{\includegraphics[width=8.5cm]{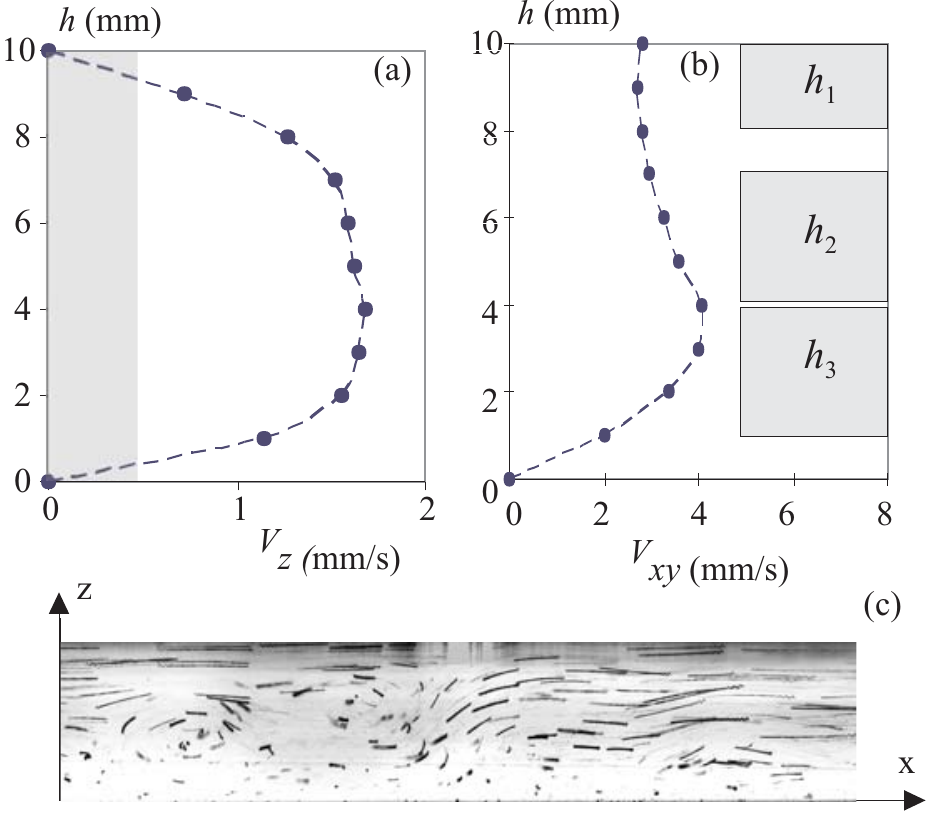}}
\caption{\label{fig2} Vertical profiles of (a) vertical, $V_z$, and (b) horizontal, $V_{x,y}$, velocities. A grey box in (a) indicates the sensitivity of the defocusing PIV technique. (c) A snapshot of the particle streaks taken at the exposure time of 2 s.}
\end{figure}

We first describe turbulent flows in a single layer. The flow is forced at $k_f \approx 800$ rad$\cdot$m$^{-1}$ ($l_f \approx 7.8$ mm) in a layer of thickness $h_l = 10$ mm. According to numerical simulations, turbulence in such a layer should show substantial three-dimensionality \cite{Celani2010}, even when the forcing is 2D. Figure \ref{fig2} shows vertical profiles of vertical (a) and horizontal (b) velocities along with the snapshot of the particle streaks in the vertical ($z-x$) plane. RMS vertical velocities $<V_z>$ are low in the top sublayer, 2 mm below the free surface, as well as in the bottom boundary sublayer. In the bulk of the flow (2 - 8 mm) $<V_z>$ is high, being only a factor of two lower than horizontal velocities $<V_{x,y}>$.

$<V_{x,y}>$ shows a maximum at $h$ = (3 - 4) mm, which is indicative of the competition between the forcing and the bottom drag. The forcing $f \sim (\mathbf{J} \times \mathbf{B})$ is the strongest near the bottom (magnets underneath the cell, the current density $J$ is constant through the layer) and it decreases inversely proportional to the distance from the bottom, $f \propto 1/h$. The bottom drag is also the strongest near the bottom; in a quasi-2D flow it should scale faster \cite{Dolzhanskii1992}, $\alpha \propto 1/h^2$, resulting in the maximum of $<V_{x,y}>$ at $h = (3 -4)$ mm. In the top sublayer, $h_1 = (8-10)$ mm, turbulence is expected to behave as quasi-2D due to the lower vertical velocities and the absence of vertical gradients of the horizontal velocities. The planarity of the top sublayer can be seen qualitatively in the particle streaks of Fig.~\ref{fig2}(c).

To test if the nature of the turbulent energy transfer changes between the top sublayer, the bulk flow and the bottom layer, we perform PIV measurements of the horizontal velocities by illuminating three different ranges of heights: $h_1 = (8 - 10)$ mm, $h_2 = (4 - 7)$ mm, and $h_2 = (1 - 4)$ mm. The velocity fields are analyzed as described, for example, in Ref.~\cite{Xia2009}. We compute the wave number energy spectra $E_k(k)$ and the third-order structure function $S_{3L} = \langle (\delta V_L)^3 \rangle$, where $\delta V_L$ is the increment across the distance $r$ of the velocity component parallel to $r$. The third-order structure function is related to the energy flux in $k$-space as $\epsilon = - (2/3) S_{3L}/r$. Positive $S_{3L}$ corresponds to negative $\epsilon$ and the inverse energy cascade.

\begin{figure*}
\centerline{\includegraphics[width=8.5cm]{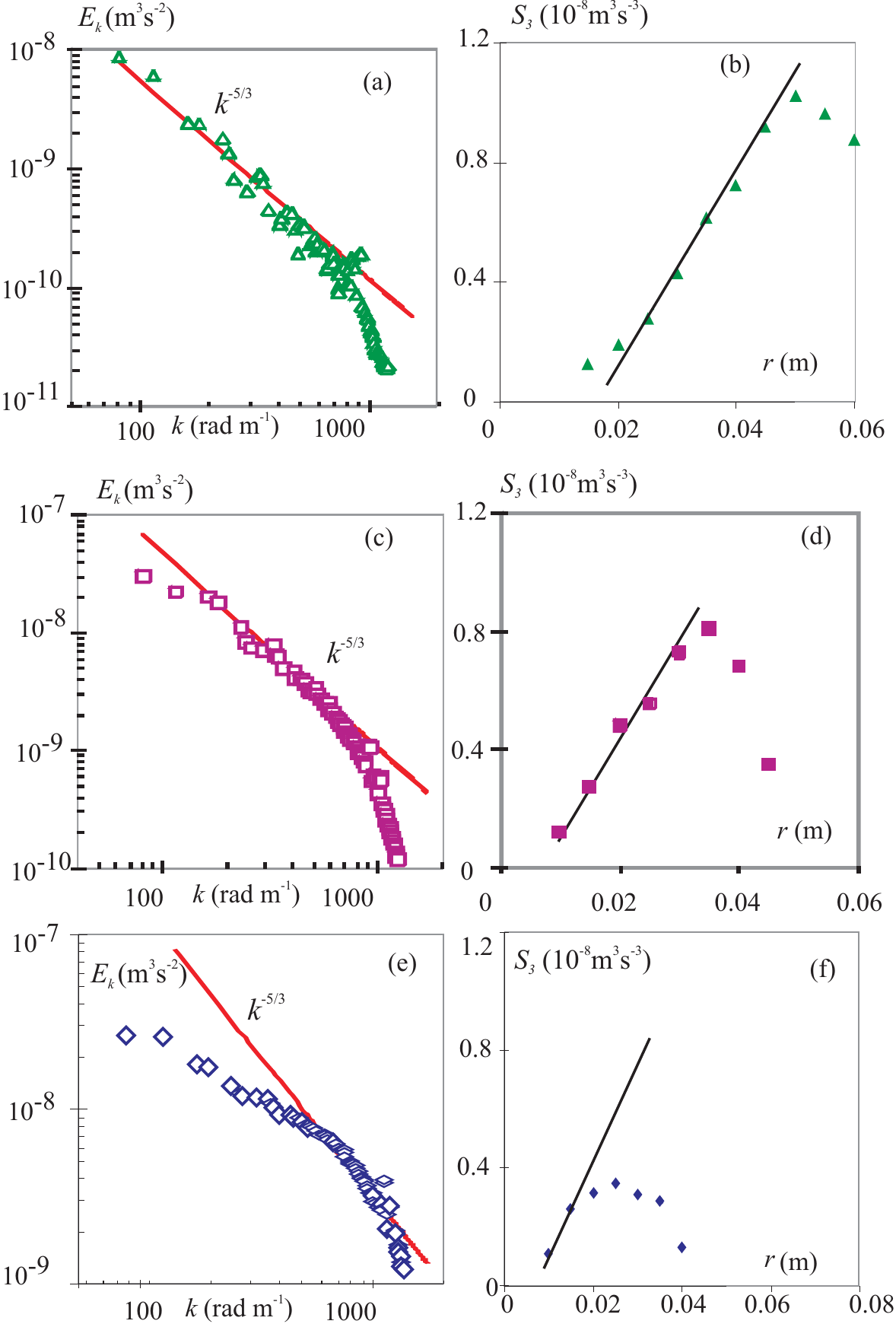}}
\caption{\label{fig3}  (a,c,e) Wave-number spectra and (b,d,f) the third order structure functions $S_{3L}$ measured in (a,b) the surface sublayer ($h_1= (8 - 10)$ mm), (c,d) in the bulk flow ($h_2= (4 - 7)$ mm), and (e,f) in the bottom sublayer regions ($h_3=(1-4)$ mm).}
\end{figure*}

Figures (\ref{fig3})(a,b) show the kinetic energy spectrum and the third-order structure function as a function of the separation distance measured in the top sublayer $h_1$. At $k < k_f$ the spectrum scales close to $k^{-5/3}$, while $S_{3L}$ is positive at $l > l_f$ and is a linear function of $r$. This is in agreement with the expectation of the quasi-2D turbulence in the top sublayer.

In the bulk flow $h_2$, which is dominated by 3D motions, the spectrum is still close to $k^{-5/3}$, though it flattens at low wave numbers as seen in Fig.\ref{fig3}(c). Consistently with this, the range of scales for which $S_{3L}$ is positive and linear is reduced to about $r \approx 40$ mm, Fig.\ref{fig3}(d). Such a behavior of $S_{3L} (r)$ is also typical for thin single layers. The reduction in the inverse energy cascade range is correlated with the increased damping. 3D motions present in the bulk flow (Fig.\ref{fig3}(c)) increase damping rate due to the increased flux of the horizontal momentum to the bottom of the cell\cite{Shats2010}. The increased damping arrests the inverse energy cascade at some scale smaller that the box size.

In the bottom sublayer $h_3$, the flow is subject to even stronger damping. As a result, the spectrum is much flatter than $k^{-5/3}$. $S_{3L}$ is positive for a narrow range of scales, giving a hint of same trend as in the bulk and the top sublayers.

The above results suggest that despite the presence of substantial 3D motion in a thick ($h/l_f = 1.28$) single layer, statistics of the horizontal velocity fluctuations remain consistent with that of quasi-2D turbulence and supports the inverse energy cascade. As seen from the energy spectra of Fig.\ref{fig3}(a,c,e), there is no evidence of the direct energy cascade at $k > k_f$. The spectrum shows that $E_k$ scales much steeper than the 3D Kolmogorov scaling of $k^{-5/3}$. A possible reason for the absence of the direct energy cascade in a 3D flow at $k > k_f$ is the fact that the Reynolds number is low in these experiments ($Re < 300$).

We performed experiments in even thicker layers, up to $h/l_f \approx 2.3$. If the layer thickness is further increased in a bounded flow, a spectral condensate forms in the top sublayer $h_1$. In a layer of a total thickness of 20 mm the spectral condensate penetrates down to 4-5 mm below the free surface. The formation of the condensate in thick layers is due to the reduction in the bottom damping, since even in the presence of 3D motions the damping rate is reduced with the increase in thickness as\cite{Shats2010} $\alpha \sim 1/h$.

\begin{figure*}[t]
\centerline{\includegraphics[width=8.5cm]{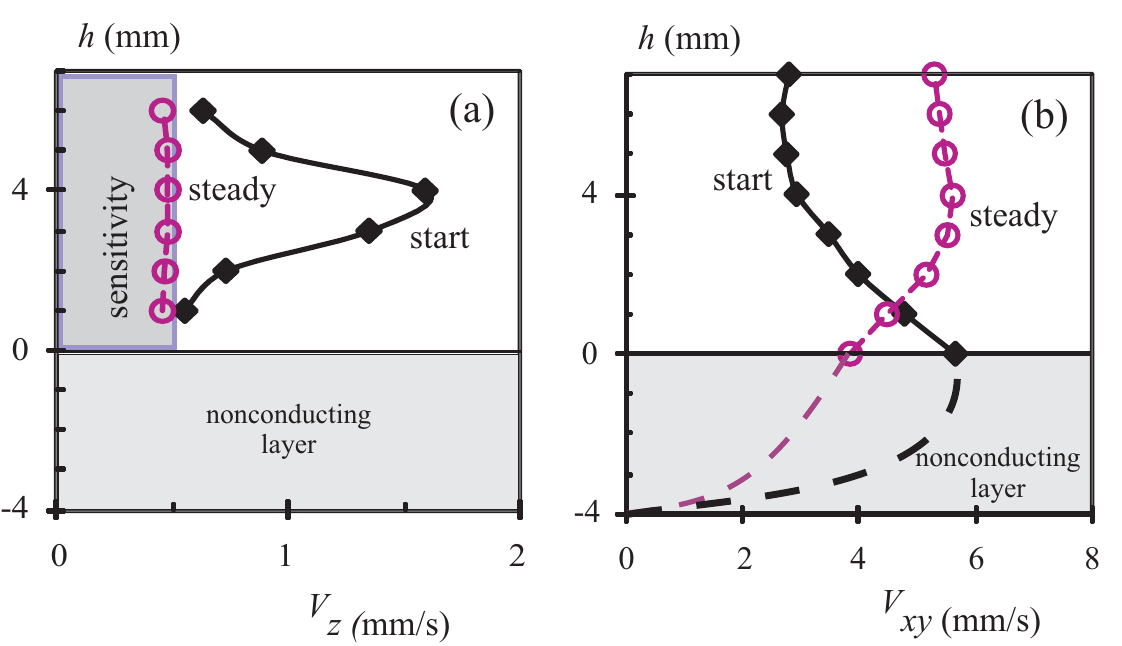}}
\caption{\label{fig4} Vertical profiles of (a) vertical, and (b) horizontal RMS velocities measured at $t =$ 5 s after forcing start, solid diamonds, and at $t =$ 20 s, open circles.}
\end{figure*}

\section{Turbulence structure in a double layer flow}

As shown in the previous section the inverse energy cascade is sustained in the presence of 3D motions. In bounded turbulence at low damping such a cascade leads to spectral condensation and to the generation of a large-scale coherent structure. Such a structure could then impose two-dimensionality on the flow in the layer, as has been found in recent experiments \cite{Xia2011}. In this section we study spatial structure of the flow during spectral condensation in thick layers subject to even lower bottom drag.

The bottom drag is reduced by generating two immiscible layers in which the bottom layer (heavier, non-conducting liquid) isolates the conduction layer from the bottom. We keep the bottom layer relatively thin, $h_b/l_f < 0.5$. The top layer, on the other hand, is thick, $h_t/l_f > 0.5$, to allow three-dimensionality to develop, even with 2D forcing. Here we study the layer configuration described in Ref.\cite{Xia2011}, namely $h_t = 7$ mm and $h_b = 4$ mm, which correspond to $h_b/l_f = 0.44$ and $h_t/l_f = 0.78$ respectively. It has been reported that the flow in the top layer shows substantial 3D motions shortly after turbulence is forced. However in the steady state, the development of spectral condensate leads to a substantial reduction of 3D eddies and to a planarization of the flow.

Figure \ref{fig4} shows vertical profiles of (a) vertical, and (b) horizontal RMS velocities measured using defocusing PIV in the top layer. The grey box in Fig.\ref{fig4}(a) indicates the method sensitivity. Shortly after the flow is forced ($t=$5 s), vertical velocities peak in the middle of the layer at  $\langle V_z \rangle \approx$ 1.7 mm/s, solid diamonds in Fig.\ref{fig4}(a). 20 seconds later vertical velocity fluctuations are substantially reduced, to below the resolution level, $\langle V_z \rangle \leq$ 0.5 mm/s. Horizontal velocities during the initial stage of the flow development peak at the interface between the two layers, i.e. in the region of strongest forcing. In the steady state, after the large coherent vortex develops the horizontal velocity profile becomes flat over most of the top layer, showing $\langle V_{x,y} \rangle \approx$ 6 mm/s at $h = 2 - 7$ mm above the interface, open circles in Fig.\ref{fig4}(b). A reduction in $\langle V_{x,y} \rangle$ near the interface in the steady state is probably related to the effect of sweeping of the forcing scale vortices by the developing condensate \cite{Shats2007}. Thus, a substantial fraction of the top layer is quasi-2D, i.e. $\langle V_z \rangle \approx$ 0 and $\partial \langle V_{x,y} \rangle / \partial z \approx$ 0.

\begin{figure}
\centerline{\includegraphics[width=8cm]{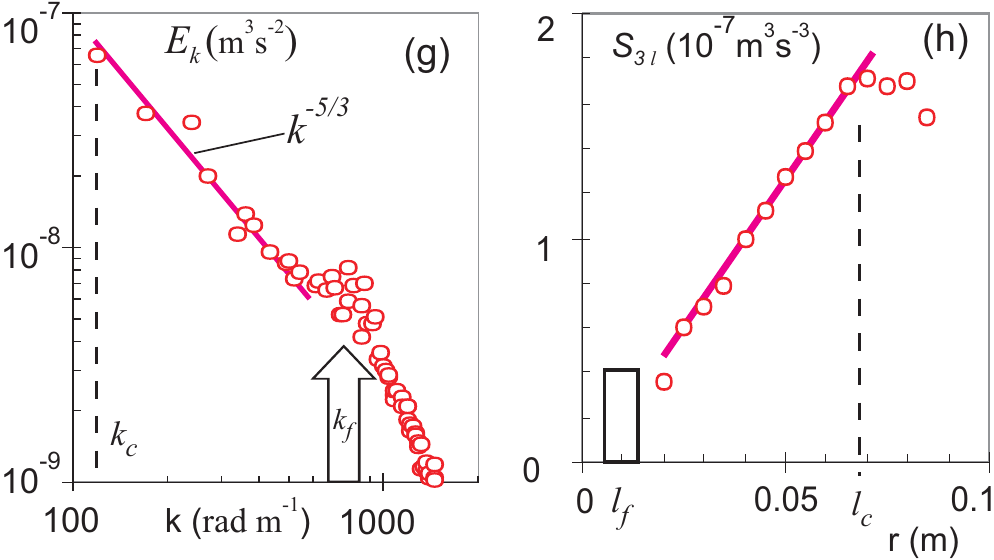}}
\caption{\label{fig5} Statistics of horizontal velocities in a double layer configuration: top layer thickness $h_t = 7$ mm, bottom layer thickness $h_t = 4$ mm. The forcing scale is $l_f = 7.8$ mm. (a) Wave number spectrum of the kinetic velocity $E_k$, and (b) the third-order structure function $S_{3L}$, computed after subtracting time-averaged mean velocity field. The entire top layer is illuminated.}
\end{figure}

Such a quasi-two-dimensionality of the flow is attributed to the shearing of vertical eddies by a strong condensate \cite{Xia2011}. Indeed, in the steady state, the statistics of horizontal velocities is in a good agreement with the Kraichnan theory. The spectra and the third-order structure functions in the presence of spectral condensate are computed after subtracting the time-average velocity field from the instantaneous velocity field, as discussed in \cite{Xia2008, Xia2009, Xia2011}. After the mean subtraction, the spectrum shows $E_k \sim k^{-5/3}$, as seen in Fig.\ref{fig5}(a). The third-order structure function is positive and is a linear function of the separation distance $r$ up to $r \approx 70$ mm, Fig.\ref{fig5}(b). Thus the spectra and the structure functions in such a flow agree with quasi-2D expectations and are consistent with the vertical structure of the flow of Fig. \ref{fig4}.

\section{Summary}

We have studied spatial structure of turbulent flows in thick layers at low Reynolds numbers ($Re \sim 100$). If the free surface of a layer is unperturbed, there is a finite thickness layer close to the surface, which remains quasi-2D regardless of the total layer thickness. Two-dimensional turbulent velocities in the top layer show kinetic energy spectra and the third-order structure functions consistent with the Kraichnan theory of 2D turbulence. Vertical eddies which appear when the layer is sufficiently thick, $h/l_f > 0.5$, introduce additional bottom drag due to the eddy viscosity \cite{Shats2010}, but they do not qualitatively change the statistics of the horizontal velocity fluctuations, which remains quasi-2D even in the presence of 3D motions. If the bottom drag is reduced by introducing an immiscible thin bottom layer, the inverse energy cascade leads to spectral condensation and to the formation of the large scale coherent structures. Such flows, as has recently been shown \cite{Xia2011}, shear off eddies in the vertical plane and reinforce quasi-two-dimensionality of the flow. Measurements presented here, in particular Fig.\ref{fig4}, confirm this.

\begin{figure}
\centerline{\includegraphics[width=8cm]{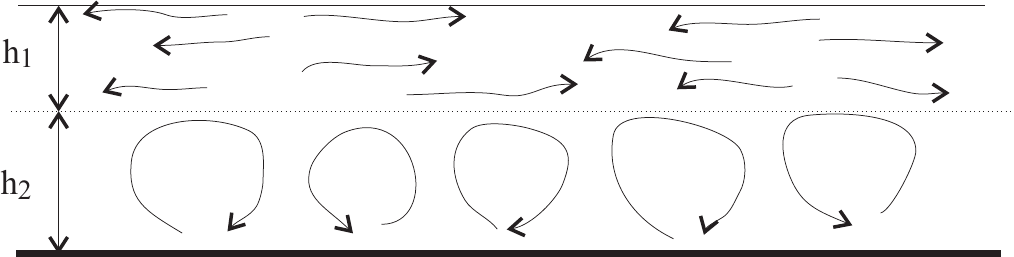}}
\caption{\label{fig6} Schematics of the structure of turbulence in thick layers. }
\end{figure}

Summarizing, flows in thick layers of fluids with an unperturbed free surface can be viewed as two interacting sublayers, as illustrated in Fig.\ref{fig6}. The top layer is quasi-2D; it supports the inverse energy cascade. In a bounded domain at low damping, the inverse cascade leads to spectral condensation of turbulence. The bottom sublayer is dominated by 3D motions which are responsible for the onset of the eddy viscosity. A planar coherent flow (spectral condensate) developing in the top layer can reduce the bottom layer thickness $h_2$ through shearing of the 3D eddies. The thickness of the two sublayers thus depends on the competition between the vertical shear and the 3D motions due to the forcing. In the two layer configuration, the spectral condensate formed in the top sublayer can take over almost the entire layer thickness.

\acknowledgments
The authors are grateful to G. Falkovich, H. Punzmann and A. Babanin for useful discussions. This work was supported by the Australian Research Council's Discovery Projects funding scheme (DP0881544). This research was supported in part by the National Science Foundation under Grant No. NSF PHY05-51164.


\begin{thebibliography}{01}

\bibitem{Kraichnan1967} R. Kraichnan, "Inertial ranges in two-dimensional turbulence", Phys. Fluids \textbf{10}, 1417 {1967}.
\bibitem{Sommeria1986} J. Sommeria, "Experimental study of the two-dimensional inverse energy cascade in a square box,"J. Fluid Mech. \textbf{170}, 139 (1986).
\bibitem{Paret_Tabeling1998} J. Paret and P. Tabeling, "Intermittency in the two-dimensional inverse cascade of energy: Experimental observations," Phys. Fluids \textbf{10}, 3126 (1998).
\bibitem{Belmonte1999} A. Belmonte, W. I. Goldburg, H. Kellay, M. Rutgers, B. Martin, and X-L. Wu, "Velocity fluctuations in a turbulent soap film: the third moment in two dimensions", Phys. Fluids \textbf{11}, 1196 (1999).
\bibitem{Vorobieff1999} P. Vorobieff, M. Rivera, and R.E. Ecke, "Soap film flows: statistics of two-dimensional turbulence", Phys. Fluids \textbf{11}, 2167 (1999).
\bibitem{Chen2006} S. Chen, R.E. Ecke, G.L. Eyink, M. Rivera, M. Wan,and M. Xiao, "Physical mechanism of the two-dimensional inverse energy cascade," Phys. Rev. Lett. \textbf{96}, 084502 (2006).
\bibitem{Shats2005} M.G. Shats, H. Xia and H. Punzmann, "Spectral condensation of turbulence in plasmas and fluids and its role in low-to-high phase transitions in toroidal plasma", Phys. Rev. E \textbf{71}, 046409 (2005).
\bibitem{Xia2008} H. Xia, H. Punzmann, G. Falkovich and M.G. Shats, "Turbulence-condensate interaction in two dimensions", Phys. Rev. Lett. \textbf{101}, 194504 (2008).
\bibitem{Xia2009} H. Xia, M. Shats and G. Falkovich, "Spectrally Condensed Turbulence in Thin Layers", Phys. Fluids \textbf{21}, 125101 (2009).
\bibitem{Xia2011} H. Xia, D. Byrne, G. Falkovich and M. Shats, "Upscale energy transfer in thick turbulent fluid layers", Nature Physics \textbf{7}, 321- 324 (2011).
\bibitem{Shats2007} M.G. Shats, H. Xia and H. Punzmann, "Suppression of turbulence by self-generated and imposed mean flows", Phys. Rev. Lett. \textbf{99}, 164502 (2007).
\bibitem{Merrifield2010} S.T. Merrifield, D.H. Kelley, and N.T. Ouellette, "Scale-dependent statistical geometry in two-dimensional flow", Phys. Rev. Lett. \textbf{104}, 254501 (2010).
\bibitem{Celani2010} A. Celani, S. Musacchio, and D. Vincenzi, "Turbulence in more than two and less than three dimensions", Phys. Rev. Lett. \textbf{104}, 184506 (2010).
\bibitem{Shats2010} M. Shats, D. Byrne and H. Xia, "Turbulence decay rate as a measure of flow dimensionality", Phys. Rev. Lett. \textbf{105}, 264501 (2010).
\bibitem{Boffetta_EPL_2005} G. Boffetta, A. Cenedese, S. Espa and S. Musacchio, "Effects of friction on 2D turbulence: An experimental study of the direct cascade", EPL \textbf{71}, 590 (2005).
\bibitem{Willert1992} C.E. Willert and M. Gharib, "Three-dimensional particle imaging with a single camera", Experiments in Fluids, \textbf {12}, 353-358 (1992).
\bibitem{Dolzhanskii1992} F.V. Dolzhanskii, V.A. Krymov and D.Yu. Manin, "An advanced experimental investigation of
quasi-two-dimensional shear flows" J. Fluid Mech. \textbf{241}, 705 (1992).



\end{thebibliography}
\end{document}